 \documentclass[final,11pt, authoryear,twocolumn]{elsarticle}

\usepackage{flushend}
\usepackage{graphicx,color,xcolor}
\usepackage{amsmath}
\usepackage{amssymb}
\usepackage{psfrag}
\usepackage{empheq}
\usepackage{bm}
\usepackage{tikz}
\usepackage{pgfplots}
\usepackage{soul}
\pgfplotsset{compat=1.17}

\usepackage{upgreek}

\newcommand{\myline}[1]{{\color{#1}{${\rule[2pt]{.5cm}{1pt}}$}}}
\newcommand{\mydash}[1]{{\color{#1}{${\rule[2pt]{.2cm}{1pt}\;\rule[2pt]{.2cm}{1pt}}$}}}

\newcommand{\thindot}[1]{{\color{#1}{${\rule[2pt]{0.05cm}{0.5pt}\,\rule[2pt]{0.05cm}{0.5pt}\,\rule[2pt]{0.05cm}{0.5pt}}$}}}

\journal{International Journal of Heat and Fluid Flow}
\begin{document}
\begin{frontmatter}



\title{Logarithmic scaling of higher-order temperature moments in the atmospheric surface layer}


\author[inst1]{Kelly Y. Huang\corref{cor1}}
\cortext[cor1]{yhuang28@nd.edu}
\author[inst2]{Matt K. Fu}
\author[inst3]{Clayton P. Byers}
\author[inst4]{Andrew D. Bragg}
\author[inst4]{Gabriel G. Katul}

\affiliation[inst1]{organization={Dept of Civil and Environmental Engineering},
            addressline={University of Notre Dame}, 
            city={Notre Dame},
            postcode={46556}, 
            state={IN},
            country={USA}}

\affiliation[inst2]{organization={GALCIT},
            addressline={Caltech}, 
            city={Pasadena},
            postcode={91125}, 
            state={CA},
            country={USA}}

\affiliation[inst3]{organization={Dept of  Engineering},
            addressline={Trinity College}, 
            city={Hartford},
            postcode={06109}, 
            state={CT},
            country={USA}}

\affiliation[inst4]{organization={Dept of Civil and Environmental Engineering},
            addressline={Duke University}, 
            city={Durham},
            postcode={27708}, 
            state={IN},
            country={USA}}

\begin{abstract}

A generalized logarithmic law for high-order moments of passive scalars is proposed for turbulent boundary layers. This law is analogous to the generalized log law that has been proposed for high-order moments of the turbulent longitudinal velocity and is derived by combining the random sweeping decorrelation hypothesis with a spectral model informed by the attached eddy hypothesis. The proposed theory predicts that the high-order moments of passive scalar fluctuations within the inertial sublayer will vary logarithmically with wall-normal distance ($z$). The proposed theory is evaluated using high frequency time-series measurements of temperature and streamwise velocity fluctuations obtained in the first meter of the atmospheric surface layer (ASL) under near-neutral thermal stratification. The logarithmic dependence with $z$ within the inertial sublayer is observed in both the air temperature and velocity moments, with good agreement to the predictions from the proposed theory. Surprisingly, the proposed theory appears to be as, if not more, valid for transported passive scalars than for the longitudinal velocity. 
\end{abstract}

\begin{graphicalabstract}
\input{figures/fig files/graphicabstract}
\end{graphicalabstract}

\begin{highlights}
\item Generalized logarithmic law for high-order moments of passive scalars derived
\item Predictions agree with experiments in close proximity to the land surface 
\end{highlights}


\end{frontmatter}



\section*{BACKGROUND}

The random sweeping decorrelation hypothesis (RSDH) is a kinematic model predicated on the assumption that small-scale eddies are swept by the larger, energetic eddies without any dynamic distortion \citep{tennekes1975eulerian}. Hence, motion of the smaller scales, particularly in the inertial range, are excited by the kinetic energy of the large scales. Given that the statistics of the large scales are often reasonably described as being Gaussian, a consequence of this kinematic assumption is the tendency towards Gaussian statistics for any flow quantities that are directly impacted by this sweeping effect. The RSDH therefore leads to significant simplifications in modeling since Gaussian random variables have known analytical expressions for their probability distribution functions and associated statistics.
In stationary and planar-homogeneous turbulent boundary layer flow at high Reynolds number ($Re$), this Gaussian behavior provides a convenient ansatz upon which generalized statistical predictions can be formed. By assuming a Gaussian probability distribution for the streamwise velocity fluctuations $u'$ and that the turbulent eddies are non-interacting, \cite{meneveau2013generalized} were able to develop a generalized logarithmic-law relation for the high-order statistics of $u'$ in the inertial sublayer given by
\begin{equation}
    \left(\overline{u'^{2p}}^+\right)^{1/p} = B_p-A_p \ln\left(\frac{z}{\delta}\right),
    \label{eq:gen_log_law_2}
\end{equation}
where $p\geq1$ is the moment order, $B_p$ is a flow-dependent constant, $A_p = A_1\left[\left(2p-1\right)!!\right]^{1/p}$, $A_1\approx 1.25$ is the Townsend-Perry constant \citep[see][]{smits2011high,marusic2013logarithmic,meneveau2013generalized}, $!!$ is the double factorial, $z$ is the wall-normal distance, $\delta$ is an outer length scale such as the boundary layer height, and the $+$ superscript indicates normalization by the friction velocity $u_*$. Following convention, $w'$ and $u'$ specify the wall-normal velocity and streamwise velocity, respectively, and primed quantities are turbulent fluctuations from the mean state that is indicated by overline. For stationary and planar homogeneous high $Re$ flow in the absence of subsidence or mean pressure gradients, the friction velocity $u_*$ is approximately equal to the shear Reynolds stress  $\sqrt{-\overline{u'w'}}$ which is independent of $z$ in the constant stress layer.

The scaling law of equation \ref{eq:gen_log_law_2} can be recovered through simpler arguments that take advantage of two hypotheses developed independently for different ranges of scales: (i) RSDH describing the behavior of small-scale motion, and (ii) the attached eddy hypothesis describing the wall effects on large-scale motion.  As shown here, the RSDH can be used to generalize the behavior of high-order spectra and the consequences of the attached eddy model can be used to describe scaling laws of the low-wavenumber properties of $u'$ \citep{katul2016generalized}.  In arriving at equation \ref{eq:gen_log_law_2}, high-order spectra were related via RSDH to a model spectrum for $u'$ characterized by a $k^{-1}$ scaling regime of wavenumber $k$ at large scales (i.e., attached eddies) and $k^{-5/3}$ for inertial scales (i.e., detached eddies) up to the Kolmogorov micro-scale. Integration of this model spectrum and its combination with RSDH then recovered equation \ref{eq:gen_log_law_2} and linked $A_p$ and $B_p$ explicitly to the Kolmogorov constant and $Re$ for Gaussian statistics.  

While RSDH has been investigated extensively in terms of the turbulent velocity, particularly $u'$, significantly less attention has been paid to the behavior of passive scalars, such as temperature in near-neutral stratification. Previous numerical studies have indicated that in isotropic turbulence, passive scalar fields adhere to the RSDH as well as, if not better than, the velocity field \citep{Yeung2002} with near-Gaussian probability distributions \citep{Overholt1996}. Further, numerous studies have empirically established the statistical similarities between streamwise velocity and passive scalar fluctuations in wall-bounded turbulence, beginning with the predictions of Reynolds analogy and more rigorous derivations of logarithmic behavior of the mean velocity and scalar profiles consistent with classical overlap arguments \citep{Kader1981,marusic2013logarithmic}. 

The similarity of these fluctuating quantities has become increasingly evident with the advent of direct numerical simulations (DNS) of turbulent channel flow, beginning with \cite{Kim1987} who found correlation coefficients between the streamwise velocity and scalar fluctuations up to $0.95$, at $Re_\tau = \delta u_*/\nu=180$, where $\delta$, here, is the channel half-height. Recent simulations at higher $Re_\tau$ up to $4088$ for $Pr = 0.7-1$ by \cite{pirozzoli2016passive} similarly revealed that the streamwise velocity variance and passive scalar variance continue to exhibit qualitatively similar wall-normal profiles. These observations were formalized by \citet{Yang2018}, who proposed a model by which passive scalars fields were interpreted through the established theoretical frameworks based on Townsend's Attached Eddy Hypothesis \citep{townsend1980structure} for streamwise velocity fluctuations. The treatment of the passive scalar field through this superposition model of wall-attached eddies explicitly recovers a $k^{-1}$ scaling behavior for wavenumbers corresponding to the attached eddies. These treatments highlight the similarities between the velocity and passive scalar field, mainly through the mechanism of the kinematics of scalars being carried by the momentum. However, care must be taken in the theoretical development for passive scalar fields. Recently, \cite{cheng2021logarithmic} investigated unstably stratified atmospheric boundary layers and how the buoyancy affects the statistics of passive scalars, determining that logarithmic behavior persisted in the mean, but with a modified slope. Their findings emphasize the need to account for non-universal behavior with the presence of thermal stratification, whereas many previous passive scalar studies assume a universal logarithmic behavior \citep{kader1972heat,Kader1981,kasagi1992direct}. In finding analogies between velocity fluctuations and temperature fluctuations, the assumption of a passive scalar field is critical. Nevertheless, even in this case, small-scale statistics of passive scalar fields can differ profoundly from those of the velocity field \cite{warhaft00}, and modeling approximations that work well for the velocity field may not work well for the scalar field (e.g. the assumption of small-scale isotropy \cite{sreenivasan1991local,warhaft00}).

Here, a new theory based on RSDH and the similarity between temperature and streamwise momentum for the behavior of passive scalar moments in the inertial sublayer is developed and tested at very high $Re$ not attainable in laboratory settings or direct numerical simulations. Following the methodology of \cite{huang2021higher}, an analogous logarithmic law for the high-order statistics of temperature fluctuations $\overline{\theta'^{2p}}$ is proposed by combining the predictions of RSDH with those from the attached eddy hypothesis in near-neutral stability conditions. The predicted behavior is found to be in agreement with highly resolved measurements of temperature obtained in the atmospheric surface layer under near-neutral conditions. Further, the deviations between the temperature statistics and their corresponding theoretical predictions are found to be comparable to and, in some instances, smaller than those found for the collocated streamwise velocity.

\section*{THEORY} 
\label{sec:thepry}
 Decades of studies have found a $k^{-1}$ scaling of temperature spectra in the atmospheric surface layer with good agreement in multiple stability conditions \citep{pond1966spectra,katul1995low,li2016,Yang2018} in addition to the measurements presented here. Combining these results with the $k^{-5/3}$ inertial range behavior through the method of \cite{katul2016generalized} and \cite{huang2021higher} allows for the theoretical development of a logarithmic law in the high-order even moments of the temperature variance given by
\begin{equation}
    \left(\overline{\theta'^{2p}}^+\right)^{1/p} = B_{p,\theta}-A_{p,\theta} \ln\left(\frac{z}{\delta}\right),
    \label{eq:gen_log_law_temp_2}
\end{equation}
where $\theta'$ is the virtual potential temperature fluctuations, $B_{p,\theta}$ is a flow-dependent offset, $A_{p,\theta}$ is the slope, analogous to the Townsend-Perry constant in equation \ref{eq:gen_log_law_2}, and the $+$ superscript again indicates normalization by the inner temperature scale $\theta_*$ given by the scalar wall flux and friction velocity, which in the constant stress layer may be approximated as $\theta_*\approx-\overline{w'\theta'}/u_*$.

 \begin{figure}[ht!]
	\centering
%
%
\begin{tikzpicture}

\begin{axis}[%
width=2.5in,
height=2.20in,
at={(0.in,0.in)},
scale only axis,
clip=false,
xmode=log,
xmin=0.0001,
xmax=10000,
xtick={\empty},
xminorticks=true,
xlabel style={font=\color{white!15!black}, align=center},
xlabel={$k$},
xtick={0.01, 1, 500},
xticklabels = {$1/\delta$, $1/z$, $1/\lambda_B$},
ymode=log,
ymin=1e-06,
ymax=10000,
ytick={\empty},
yminorticks=true,
ylabel style={font=\color{white!15!black}},
ylabel={$E_\theta(k)$},
axis background/.style={fill=white},
axis x line*=bottom,
axis y line*=left
]
\addplot [color=black, dashdotted, forget plot]
  table[row sep=crcr]{%
0.0001	0.01\\
0.0002	0.04\\
0.0003	0.09\\
0.0004	0.16\\
0.0005	0.25\\
0.0006	0.36\\
0.0007	0.49\\
0.0008	0.64\\
0.0009	0.81\\
0.001	1\\
0.0011	1.21\\
0.0012	1.44\\
0.0013	1.69\\
0.0014	1.96\\
0.0015	2.25\\
0.0016	2.56\\
0.0017	2.89\\
0.0018	3.24\\
0.0019	3.61\\
0.002	4\\
0.0021	4.41\\
0.0022	4.84\\
0.0023	5.29\\
0.0024	5.76\\
0.0025	6.25\\
0.0026	6.76\\
0.0027	7.29\\
0.0028	7.84\\
0.0029	8.41\\
0.003	9\\
0.0031	9.61\\
0.0032	10.24\\
0.0033	10.89\\
0.0034	11.56\\
0.0035	12.25\\
0.0036	12.96\\
0.0037	13.69\\
0.0038	14.44\\
0.0039	15.21\\
0.004	16\\
0.0041	16.81\\
0.0042	17.64\\
0.0043	18.49\\
0.0044	19.36\\
0.0045	20.25\\
0.0046	21.16\\
0.0047	22.09\\
0.0048	23.04\\
0.0049	24.01\\
0.005	25\\
0.0051	26.01\\
0.0052	27.04\\
0.0053	28.09\\
0.0054	29.16\\
0.0055	30.25\\
0.0056	31.36\\
0.0057	32.49\\
0.0058	33.64\\
0.0059	34.81\\
0.006	36\\
0.0061	37.21\\
0.0062	38.44\\
0.0063	39.69\\
0.0064	40.96\\
0.0065	42.25\\
0.0066	43.56\\
0.0067	44.89\\
0.0068	46.24\\
0.0069	47.61\\
0.007	49\\
0.0071	50.41\\
0.0072	51.84\\
0.0073	53.29\\
0.0074	54.76\\
0.0075	56.25\\
0.0076	57.76\\
0.0077	59.29\\
0.0078	60.84\\
0.0079	62.41\\
0.008	64\\
0.0081	65.61\\
0.0082	67.24\\
0.0083	68.89\\
0.0084	70.56\\
0.0085	72.25\\
0.0086	73.96\\
0.0087	75.69\\
0.0088	77.44\\
0.0089	79.21\\
0.009	81\\
0.0091	82.81\\
0.0092	84.64\\
0.0093	86.49\\
0.0094	88.36\\
0.0095	90.25\\
0.0096	92.16\\
0.0097	94.09\\
0.0098	96.04\\
0.0099	98.01\\
0.01	100\\
};
\addplot [color=black, forget plot]
  table[row sep=crcr]{%
0.01	100\\
0.02	50\\
0.03	33.3333333333333\\
0.04	25\\
0.05	20\\
0.06	16.6666666666667\\
0.07	14.2857142857143\\
0.08	12.5\\
0.09	11.1111111111111\\
0.1	10\\
0.11	9.09090909090909\\
0.12	8.33333333333333\\
0.13	7.69230769230769\\
0.14	7.14285714285714\\
0.15	6.66666666666667\\
0.16	6.25\\
0.17	5.88235294117647\\
0.18	5.55555555555556\\
0.19	5.26315789473684\\
0.2	5\\
0.21	4.76190476190476\\
0.22	4.54545454545455\\
0.23	4.34782608695652\\
0.24	4.16666666666667\\
0.25	4\\
0.26	3.84615384615385\\
0.27	3.7037037037037\\
0.28	3.57142857142857\\
0.29	3.44827586206897\\
0.3	3.33333333333333\\
0.31	3.2258064516129\\
0.32	3.125\\
0.33	3.03030303030303\\
0.34	2.94117647058824\\
0.35	2.85714285714286\\
0.36	2.77777777777778\\
0.37	2.7027027027027\\
0.38	2.63157894736842\\
0.39	2.56410256410256\\
0.4	2.5\\
0.41	2.4390243902439\\
0.42	2.38095238095238\\
0.43	2.32558139534884\\
0.44	2.27272727272727\\
0.45	2.22222222222222\\
0.46	2.17391304347826\\
0.47	2.12765957446809\\
0.48	2.08333333333333\\
0.49	2.04081632653061\\
0.5	2\\
0.51	1.96078431372549\\
0.52	1.92307692307692\\
0.53	1.88679245283019\\
0.54	1.85185185185185\\
0.55	1.81818181818182\\
0.56	1.78571428571429\\
0.57	1.75438596491228\\
0.58	1.72413793103448\\
0.59	1.69491525423729\\
0.6	1.66666666666667\\
0.61	1.63934426229508\\
0.62	1.61290322580645\\
0.63	1.58730158730159\\
0.64	1.5625\\
0.65	1.53846153846154\\
0.66	1.51515151515152\\
0.67	1.49253731343284\\
0.68	1.47058823529412\\
0.69	1.44927536231884\\
0.7	1.42857142857143\\
0.71	1.40845070422535\\
0.72	1.38888888888889\\
0.73	1.36986301369863\\
0.74	1.35135135135135\\
0.75	1.33333333333333\\
0.76	1.31578947368421\\
0.77	1.2987012987013\\
0.78	1.28205128205128\\
0.79	1.26582278481013\\
0.8	1.25\\
0.81	1.23456790123457\\
0.82	1.21951219512195\\
0.83	1.20481927710843\\
0.84	1.19047619047619\\
0.85	1.17647058823529\\
0.86	1.16279069767442\\
0.87	1.14942528735632\\
0.88	1.13636363636364\\
0.89	1.12359550561798\\
0.9	1.11111111111111\\
0.91	1.0989010989011\\
0.92	1.08695652173913\\
0.93	1.0752688172043\\
0.94	1.06382978723404\\
0.95	1.05263157894737\\
0.96	1.04166666666667\\
0.97	1.03092783505155\\
0.98	1.02040816326531\\
0.99	1.01010101010101\\
1	1\\
};
\addplot [color=black, dashed, forget plot]
  table[row sep=crcr]{%
1	1\\
6.04040404040404	0.0499141122114033\\
11.0808080808081	0.0181571470866859\\
16.1212121212121	0.00972009506904241\\
21.1616161616162	0.006176635105938\\
26.2020202020202	0.00432622103633654\\
31.2424242424242	0.00322670150015495\\
36.2828282828283	0.00251476907479643\\
41.3232323232323	0.00202461690374267\\
46.3636363636364	0.00167123497221228\\
51.4040404040404	0.00140714193606926\\
56.4444444444444	0.00120401342269834\\
61.4848484848485	0.00104404645094875\\
66.5252525252525	0.000915564855566581\\
71.5656565656566	0.000810635402744348\\
76.6060606060606	0.000723704795559718\\
81.6464646464647	0.000650785237824137\\
86.6868686868687	0.000588949160863468\\
91.7272727272727	0.000536005440946503\\
96.7676767676768	0.000490285981505168\\
101.808080808081	0.000450501507923062\\
106.848484848485	0.000415641962307963\\
111.888888888889	0.000384906340139388\\
116.929292929293	0.000357652386647661\\
121.969696969697	0.000333359952287221\\
127.010101010101	0.000311603909343681\\
132.050505050505	0.000292033869092712\\
137.090909090909	0.000274358807190649\\
142.131313131313	0.00025833527932853\\
147.171717171717	0.000243758295683747\\
152.212121212121	0.000230454186928565\\
157.252525252525	0.0002182749778339\\
162.292929292929	0.000207093913361385\\
167.333333333333	0.00019680187386769\\
172.373737373737	0.000187304482112136\\
177.414141414141	0.000178519752861518\\
182.454545454545	0.000170376171267005\\
187.49494949495	0.000162811112460587\\
192.535353535354	0.000155769534503072\\
197.575757575758	0.000149202891688776\\
202.616161616162	0.000143068226539477\\
207.656565656566	0.000137327407512158\\
212.69696969697	0.000131946486162214\\
217.737373737374	0.000126895152729909\\
222.777777777778	0.000122146273209837\\
227.818181818182	0.000117675494186731\\
232.858585858586	0.000113460904274983\\
237.89898989899	0.000109482743034061\\
242.939393939394	0.000105723149861633\\
247.979797979798	0.000102165946677817\\
253.020202020202	9.87964492748569e-05\\
258.060606060606	9.56013030683109e-05\\
263.10101010101	9.25683396892046e-05\\
268.141414141414	8.96864514328913e-05\\
273.181818181818	8.69454810545694e-05\\
278.222222222222	8.43361247930672e-05\\
283.262626262626	8.18498468292045e-05\\
288.30303030303	7.94788036552e-05\\
293.343434343434	7.72157770571416e-05\\
298.383838383838	7.50541146014646e-05\\
303.424242424242	7.29876766751489e-05\\
308.464646464646	7.10107892631738e-05\\
313.505050505051	6.91182017599074e-05\\
318.545454545455	6.73050492070421e-05\\
323.585858585859	6.55668184322452e-05\\
328.626262626263	6.38993176322279e-05\\
333.666666666667	6.22986490033686e-05\\
338.707070707071	6.07611840739475e-05\\
343.747474747475	5.92835414357976e-05\\
348.787878787879	5.78625666108356e-05\\
353.828282828283	5.64953138204385e-05\\
358.868686868687	5.51790294537445e-05\\
363.909090909091	5.39111370553145e-05\\
368.949494949495	5.26892236737576e-05\\
373.989898989899	5.15110274313386e-05\\
379.030303030303	5.03744261906438e-05\\
384.070707070707	4.92774272084178e-05\\
389.111111111111	4.8218157678962e-05\\
394.151515151515	4.71948560802593e-05\\
399.191919191919	4.62058642454504e-05\\
404.232323232323	4.52496200906123e-05\\
409.272727272727	4.43246509371307e-05\\
414.313131313131	4.34295673734363e-05\\
419.353535353535	4.2563057606604e-05\\
424.393939393939	4.17238822593914e-05\\
429.434343434343	4.09108695727897e-05\\
434.474747474747	4.01229109781646e-05\\
439.515151515152	3.93589570066144e-05\\
444.555555555556	3.8618013506348e-05\\
449.59595959596	3.78991381417055e-05\\
454.636363636364	3.72014371499754e-05\\
459.676767676768	3.65240623344129e-05\\
464.717171717172	3.58662082738937e-05\\
469.757575757576	3.52271097314446e-05\\
474.79797979798	3.46060392455235e-05\\
479.838383838384	3.40023048893836e-05\\
484.878787878788	3.34152481851731e-05\\
489.919191919192	3.28442421606068e-05\\
494.959595959596	3.22886895371181e-05\\
500	3.1748021039364e-05\\
};
\addplot [color=black, dotted, forget plot]
  table[row sep=crcr]{%
0.01	1e-06\\
0.01	100\\
};
\addplot [color=black, dotted, forget plot]
  table[row sep=crcr]{%
1	1e-06\\
1	1\\
};
\addplot [color=black, dotted, forget plot]
  table[row sep=crcr]{%
500	1e-06\\
500	3.1748021039364e-05\\
};
\node[above, align=center]
at (axis cs:0.001,100) {$C_1\theta_*^2\delta^3k^2$};
\node[above right, align=left]
at (axis cs:0.1,10) {$C_1\theta_*^2k^{-1}$};
\node[above right, align=left]
at (axis cs:15,0.01) {$C_o\overline{\epsilon}^{-\frac{1}{3}}\overline{\epsilon}_\theta k^{-\frac{5}{3}}$};

\addplot[area legend, draw=black, fill=black, forget plot]
table[row sep=crcr] {%
x	y\\
10000	1.2e-06\\
10000	8e-07\\
15000	1e-06\\
}--cycle;

\addplot[area legend, draw=black, fill=black, forget plot]
table[row sep=crcr] {%
x	y\\
8.8e-05	10000\\
0.000112	10000\\
0.0001	18000\\
}--cycle;
\end{axis}

\end{tikzpicture}%
	\caption{Illustration of an idealized piece-wise temperature spectrum and its scaling regimes as a function of wavenumber $k$.  The $k^2$ behavior (analogous to the Saffman spectrum) is the minimum required for $dE_{\theta}/dk\rightarrow0$ as $k\rightarrow0$. For analytical tractability, the Kolmogorov $k^{-5/3}$ inertial scaling is extended to the Batchelor scale with a cutoff compensating for the energy contribution not resolved here by the expected exponential decay due to molecular effects.}
	\label{figure1}
\end{figure}
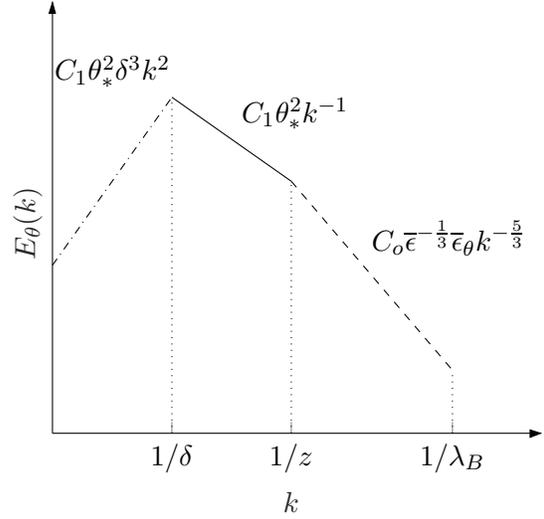

Equation \ref{eq:gen_log_law_temp_2} is derived by considering the time series for the passive scalar quantity, $\theta'$. The integral of the $p^{\mathrm{th}}$-order spectrum, $E_{\theta}^{(p)}(k)$, is related to statistical moments of $\theta'$ through
\begin{equation} 
	\overline{(\theta'^p-\overline{\theta'^p})^2}=	\overline{\theta'^{2p}}-	\overline{\theta'^p}^2=\int_{0}^{+\infty} E_{\theta}^{(p)}(k)\,dk.
	\label{eqn:variance_int}
\end{equation} 
Under the assumptions of RSDH, the probability distribution function (PDF) of the time series is taken to be normally distributed, allowing the higher-order statistical moments and spectra to be expressed analytically as a function of the second ($p=1$) moment resulting in
\begin{equation}
	E_{\theta}^{(p)}(k) = \alpha(p) \overline{\theta'^2}^{p-1}E_{\theta}(k),
\label{eqn:HO_spectra}
\end{equation}  
where $$\alpha(p)=p^2 \times 1 \times 3 \times 5 \times \ldots \times (2p-3)$$ is a consequence of Gaussian statistics for $p>1$. 

The high-order moments for temperature $$\overline{\theta'^{2p}} =\int_{0}^{\infty} E_{\theta}^{(p)}(k)\,dk + \overline{\theta'^{p}}^2 $$ are now interpreted with the RSDH (see equation \ref{eqn:HO_spectra}) to yield 
\begin{equation}
    \overline{\theta'^{2p}}=\alpha(p) \left[\overline{\theta'^2}\right]^{p-1} \int_{0}^{\infty} E_{\theta}(k)\,dk + \overline{\theta'^{p}}^2.
    \label{eqn:HO_RSD_full}
\end{equation}
It can be shown for Gaussian statistics that the $\overline{\theta'^{p}}^2$ term is zero for odd $p$ and diminishes rapidly for even $p$ so that its contribution is only significant for $p=2$ with a value of $(3/2) \overline{\theta'^{2p}}$.

To arrive at an analytic expression, an idealized shape for $E_\theta(k)$ is prescribed and shown in figure \ref{figure1} based on \cite{huang2021higher} and given by the piecewise function
\begin{equation}
E_\theta(k)=\begin{cases}
          C_1\theta_*^2\delta^3k^2 \ &\; \text{if} \, k \in [0,\delta^{-1}]\\
          C_1\theta_*^2k^{-1} \ &\; \text{if} \, k \in [\delta^{-1},z^{-1}] \\
          C_\theta  \overline{\epsilon}^{-1/3} \overline{\epsilon_\theta}k^{-5/3} \ &\;\text{if} \, k \in [z^{-1},\lambda_B^{-1}]\\
          0 \ &\;\text{if} \, k > \lambda_B^{-1},\\
     \end{cases}
     \label{eq:specral_model}
\end{equation}
\noindent where $\lambda_B=Sc_m^{-1/2} \eta$ is the Batchelor scale, close to the Kolmogorov microscale $\eta$ in near-neutral atmospheric flows because the molecular Schmidt number ($Sc_m$) is close to unity for temperature and many other scalars (and equation \eqref{eq:specral_model} already assumes $Sc=O(1)$ when prescribing the spectrum scaling for $k \in [z^{-1},\lambda_B^{-1}]$). Here, we have adopted the canonical scaling laws in the high wavenumber regime based on Kolmogorov-Obukhov-Corrsin (KOC) theory \citep{kolmogorov1941local,obukhov1949temperature,Corrsin1951} for isotropic turbulence. For the applications of interest such as the atmospheric surface layer, shear effects on spectral exponents in the inertial subrange can be assumed to be minimal  \citep{tennekes1972first}. However, in locations or applications where anisotropy effects may not be negligible, alternative scaling may need to be adopted to account for such effects, e.g., see the $-4/3$ scaling proposed by \citet{Lohse1994}. The low wavenumber behavior (i.e., $k\leq\delta^{-1}$) is modeled after \cite{Saffman1967} to ensure the energy contribution in the very large scales is not singular as $k\rightarrow 0$. This choice is a deviation from the functional form proposed by \cite{katul2016generalized}, who applied a constant value in this region. Assuming that production of turbulent kinetic energy and temperature variance are each balanced by their respective dissipation within the inertial sublayer  (i.e.,  $\overline{\epsilon}=u_*^3/\kappa z$ and $\overline{\epsilon}_\theta = u_*\theta_*^2/\kappa z$,  and $\kappa\approx0.41$), the spectral model in equation \ref{eq:specral_model} can be integrated, giving
\begin{align*}
    \int_{0}^{\infty} E_{\theta}(k)\,dk =\overline{\theta'^2}=\int_{0}^{1/\delta} C_1\theta_*^2\delta^3k^2 \, dk+ \\ \int_{1/\delta}^{1/z} C_1\theta_*^2k^{-1} \, dk + \int_{1/z}^{1/\lambda_B} C_\theta \overline{\epsilon_\theta}^{-2/3}k^{-5/3}\, dk,
\end{align*}
allowing the temperature variance to ultimately be expressed as
\begin{multline}
    \frac{\overline{\theta'^2}}{\theta_*^2} \equiv \overline{\theta'^{2}}^+ = C_1\left[\frac{11}{6}-\frac{3}{2}\left(\frac{\lambda_B}{z} \right)^{2/3}\right]\\-C_1 \text{ln}\left(\frac{z}{\delta}\right).
    \label{eqn:sigma2}
\end{multline}
 Equation \ref{eqn:sigma2} recovers the logarithmic form proposed in equation \ref{eq:gen_log_law_temp_2} where $$A_{1,\theta}=C_1$$ and $$B_{1,\theta} = C_1\left[\frac{11}{6}-\frac{3}{2}\left(\frac{\lambda_B}{z} \right)^{2/3}\right].$$ Note the value of $C_\theta$ is related to $C_1$ through matching the piecewise function at $k=1/z$. Equation \ref{eqn:HO_RSD_full} can then be written as 
\begin{equation}
    \overline{\theta'^{2p}} = \alpha(p) \left[\overline{\theta'^2}\right]^{p-1} \theta_*^2 \left[B_{1,\theta}-A_{1,\theta} \text{ln}\left(\frac{z}{\delta}\right)\right].
    \label{eqn:RSD_intermediate}
\end{equation}
Finally, substituting $$\overline{\theta'^2}=\theta_*^2 \left[ B_{1,\theta}-A_{1,\theta}\text{ln}\left(\frac{z}{\delta}\right)\right],$$ normalizing by $\theta_*^{2p}$, and raising to a power of $1/p$ provides the sought result
\begin{equation}
    (\overline{\theta'^{2p}}^+) ^{1/p} = \alpha(p)^{1/p}\left[B_{1,\theta}-A_{1,\theta}\text{ln}\left(\frac{z}{\delta}\right)\right].
    \label{eqn:HO_mom}
\end{equation}
This expression has the same form as equation \ref{eq:gen_log_law_temp_2} with $$A_{p,\theta}= \alpha(p)^{1/p}A_{1,\theta}$$ and $$B_{p,\theta}= \alpha(p)^{1/p}B_{1,\theta}.$$








\section*{EXPERIMENTAL METHODOLOGY \& DATA}
As part of the Idealized horizontal Planar Array study for Quantifying Surface heterogeneity (IPAQS) \citep{Morrison2021imapct} that took place at the Surface Layer Turbulence and Environmental Science Test (SLTEST) facility in the western deserts of Utah, USA, measurements of longitudinal velocity and temperature were acquired at $z$ = 0.0625, 0.125, 0.25, 0.5, and 1.0 m above the ground during a three-day intensive sampling period (18 - 20 June 2018). The SLTEST facility is a unique field site with near-canonical boundary conditions (its aerodynamic roughness ranges from sub-millimeter to less than 6 mm) and predictable wind patterns, making it an ideal location for probing high $Re$ turbulent boundary layer flows \citep{klewicki1998high, metzger2001comparative}. The equivalent sand grain roughness for these measurements was estimated to be 2.5 mm using the relation for a zero-pressure-gradient neutral boundary layer \citep{Huang2021}, meaning the lowest station height corresponds to 25 times that of the equivalent sand grain roughness.

\begin{table*}
\centering
\small
    \begin{tabular}{cccccc}
    \noalign{\smallskip}
  LT & $\overline{u}$ & $u_*$ & $\overline{w'\theta_v'}$ & $\zeta$ & $R_f$ \\
     & (m$\cdot$s$^{-1}$) & (m$\cdot$s$^{-1}$) & (K$\cdot$m$\cdot$s$^{-1}$)& &\\
    \noalign{\smallskip}\hline\hline\noalign{\smallskip}
  2000 & 7.55 & 0.28 & -0.017 & 0.025 & 0.014 \\
  2030 & 6.56 & 0.22 & -0.018 & 0.053 & 0.040 \\
  2200 & 6.60 & 0.24 & -0.024 & 0.056 & 0.035 \\
  2230 & 7.34 & 0.29 & -0.029 & 0.038 & 0.016 \\


    \noalign{\smallskip}\hline
    \end{tabular}
    \caption{Mean flow properties relative to the triaxial sonic anemometer positioned at $z$ = 2 m for the 30-min periods analyzed (local time LT = UTC + 6 h) on 21/06/2018. The boundary layer height $\delta$ was estimated to be 60 m \citep{huang2021velocity}, and $R_f$ was estimated using both the sonic anemometer data and the mean velocity profile given by the NSTAPs evaluated at $z=2$ m.}
    \label{tab:data_summary}       
\end{table*}

Nano-scale thermal anemometry probes (NSTAPs) operated in constant-current anemometry mode and their cold-wire variants (TNSTAPs) were used to capture velocity and temperature measurements, respectively, at each height with a sampling frequency of 100 Hz. The sensing elements of these nano-scale sensors are platinum wire ribbons 2 $\upmu$m in width, 100 nm in thickness, 60 $\upmu$m in length for the NSTAP and 200 $\upmu$m in length for the TNSTAP. The size of these sensors provides high spatial resolution with minimized end-conduction effects and temporal filtering \citep{hultmark2011endconduction,arwatz2015development}. Relevant stresses and heat fluxes are calculated from data acquired by a nearby triaxial sonic anemometer (Campbell Scientific CSAT3) at approximately 10 m west of the measurement tower at $z=$ 2 m. The approximate number of large eddy turnovers ranges from 196 to 1180, depending on the particular calculation. Note this is likely an underestimate, as the mean convective velocity $\bar{u}$ is calculated at the $2$ m height and not the freestream velocity. Further details on the different estimates of eddy turnovers can be found in the appendix. Details regarding the experimental set-up and calibration procedure can be found elsewhere \citep{Huang2021}.

Trends associated with the varying free stream velocity were subtracted from the raw time series  \citep{hutchins2012towards}. Four 30-min records are examined in the current study and summarized in Table \ref{tab:data_summary} after the following data qualifications: 
\begin{enumerate}
    \item The incoming wind direction aligned with the sensors, as verified by the sonic anemometer. 
    \item Near-neutral stability was reached, with both the Monin-Obukhov stability parameter $|\zeta|=|z/L|\approx 0$ and the flux Richardson number $R_f\approx0$. The Obukhov length $L$ is defined as 
    \begin{equation}
        L = -\frac{u_*^3}{\kappa \left(g/{\overline{\theta_v}}\right) \overline{w'\theta_v'}},
    \label{eqn:L_o_intro}
    \end{equation}
    where $g$ is the acceleration due to gravity, and $\overline{\theta_v}$ is the mean virtual potential temperature. The flux Richardson number $R_f$ is defined as
    \begin{equation}
        R_f = \frac{(g/\overline{\theta_v})\overline{w'\theta_v'}}{\overline{w'u'}(d\overline{u}/dz)}.
        \label{eqn:R_f}
    \end{equation}
    \item Stationarity of $\overline{u}$ and turbulence intensity $u'/\overline{u}$ were observed, as assessed by the reverse arrangement test and the runs test with a 95\% confidence interval \citep{bendat2011random}.

\end{enumerate}
The four runs satisfying these conditions were near mid-night (local time) when the stable boundary layer depth has equilibrated to its near-neutral value. Taylor's hypothesis was used to convert temporal differences in the velocity and temperature signals to spatial differences in the streamwise direction.

\begin{figure*}[ht!]
	\centering
\input{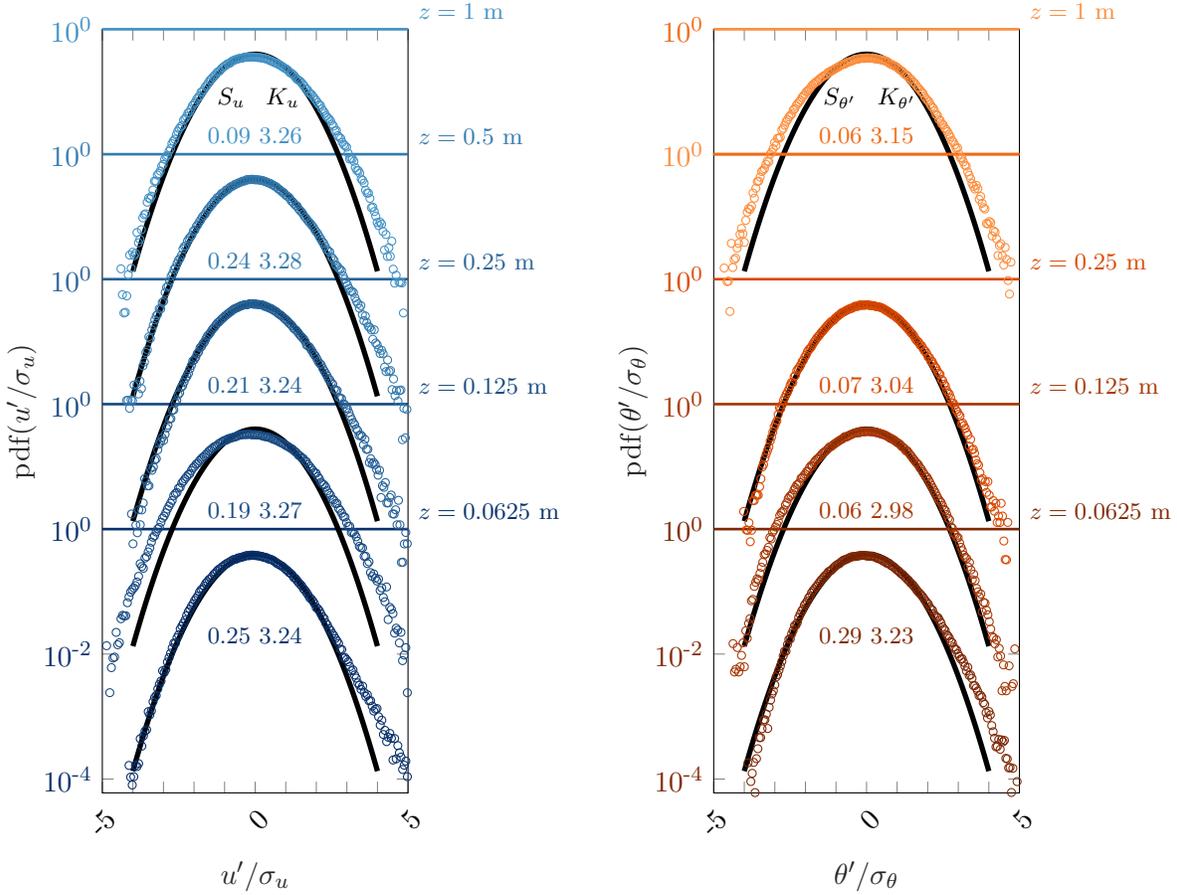}
	\caption{Measured probability density functions (open symbols) of the velocity (left) and temperature (right) fluctuations at $z=1$ m, $z=0.5$ m, $z=0.25$ m, $z=0.125$ m, and $z=0.0625$ m. The solid black line indicates a Gaussian distribution.}
	\label{figure2}
\end{figure*}

\section*{RESULTS}
Before evaluating the model predictions, this  dataset can be used to first assess the validity of the key assumptions underpinning the proposed model.  These are (i) the tendency towards Gaussian statistics, (ii) the existence of inertial range behavior, and (iii) spectral behavior consistent with the attached eddy model. The Gaussian behavior of the velocity and temperature fluctuations can be assessed using the PDFs of the measurements. The PDFs for each quantity are shown in figure \ref{figure2} at each measurement station height. The PDFs are normalized by their standard deviation $\sigma_{u}$ and $\sigma_\theta$, respectively and their skewness $S_{u'}$ and $S_{\theta'}$ and kurtosis $K_{u'}$ and $K_{\theta'}$ are shown. The PDFs of both temperature and velocity exhibit near-Gaussian behavior, with the most evident deviations being super-Gaussian in the positive tail. 

\begin{figure*}[ht!]
	\centering
\input{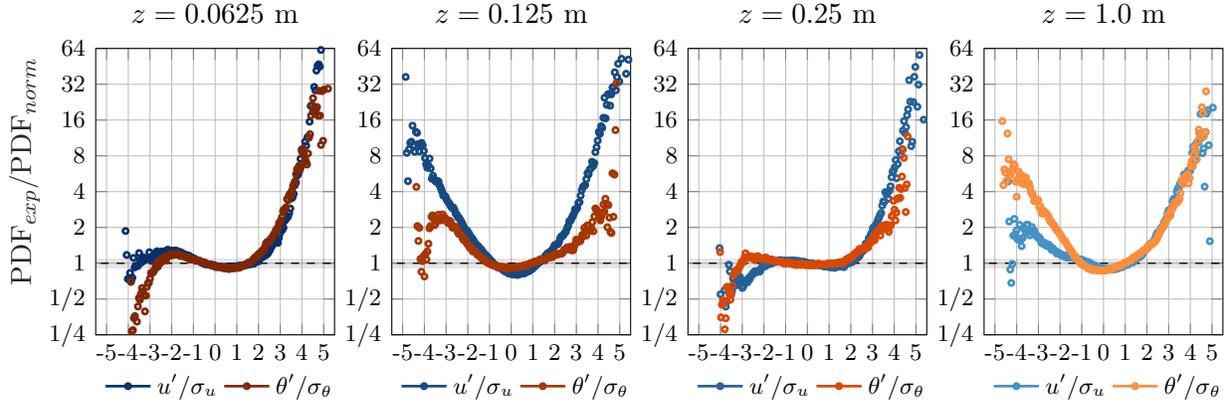}

	\caption{Deviations of measured PDFs from Gaussian behavior. PDFs of measured velocity (blue) and temperature (orange) are normalized by a Gaussian distribution given by $\mathrm{PDF}_{norm} = \exp\left(-x^2/2\right)/\sqrt{2\pi}$, where $x=u'/\sigma_{u},\theta'/\sigma_\theta$. Dashed black line (---) at unity indicates Gaussian behavior with a gray band showing the region of $\pm10\%$ deviation. Values less than and greater than 1 indicate sub-Gaussian and super-Gaussian behavior, respectively.}
	\label{figure3}
\end{figure*}

The deviations of these velocity and temperature PDFs from Gaussian behavior are highlighted in figure \ref{figure3} for each measurement height where collocated measurements were obtained. There, each of the PDFs from figure \ref{figure2} is normalized by the Gaussian distribution such that deviations from Gaussian behavior are indicated in proportion to unity. It can be seen that the scalar and velocity PDFs exhibit similar behavior for $-2<\theta'/\sigma_\theta(\text{or } u'/\sigma_u) < 2$. For $z=0.125$ m and $z=0.25$ m, the scalar PDFs are found to be within $\pm 10\%$ of a Gaussian distribution for a wider region, and exhibit smaller deviations from Gaussian behavior in the tails compared to their velocity counterparts. Though the exact behavior and statistics differ with station height, the tendency for the temperature fluctuations towards similar Gaussian behavior relative to the collocated velocity is a consistent feature across all $z$.  These findings agree with prior sonic anemometry measurements collected above a dry lake bed \cite{chu1996probability}.

The super-Gaussian behavior in the positive tails differs from the observations of \cite{meneveau2013generalized} and \cite{samie2018} who observed sub-Gaussian behavior for boundary layer measurements up to $Re_\tau=20{,}000$, and is potentially indicative of effects stemming from the roughness sublayer \citep{heisel2020velocity}. Recent studies such as \cite{marusic2013logarithmic} and \cite{wei2005properties} suggest that the extent of the buffer layer could be Reynolds number dependent, extending up to $z^+\approx 3 (\delta u_*/\nu)^{1/2}\approx 3 \times 10^3$, with $\nu$ being kinematic viscosity, even in smooth-wall conditions. This criteria places the lowest two measurement heights within the buffer region where sweeps tend to dominate \citep{heisel2020velocity}, potentially explaining the observed increase in $S_{u'}$ and $K_{u'}$.

\begin{figure}[ht!] 
	\centering
%
%
\definecolor{mycolor1}{rgb}{0.14510,0.38039,0.59804}%
\definecolor{mycolor2}{rgb}{0.85098,0.28235,0.00392}%

\newcommand\Height{1.45in}
\newcommand\Width{2in}
\newcommand\VSKIP{1.55in}

\begin{tikzpicture}

\begin{axis}[%
width=\Width,
height=\Height,
at={(0in,2.50in)},
scale only axis,
xmode=log,
xmin=200,
xmax=50000,
xtick={1000,10000},
xticklabels={{$\text{10}^\text{3}$},{$\text{10}^\text{4}$}},
xminorticks=true,
xlabel style={font=\color{white!15!black}},
xlabel={$z^{+}$},
separate axis lines,
ytick pos=left,
ylabel shift = -8pt,
ylabel style={rotate=-90},
every outer y axis line/.append style={black},
every y tick label/.append style={font=\color{black}},
every y tick/.append style={black},
ylabel={\color{black}$U^+$},
title = {Velocity Profiles},
ymin=15,
ymax=27,
axis background/.style={fill=none}
]
\addplot [color=black, line width=1pt, forget plot]
  table[row sep=crcr]{%
1070.20833333333	17.5787232004561\\
17123.3083333333	25.2160746782899\\
};
\addplot [color=black, line width=0pt, draw=none, mark=square*, mark size=2pt,mark options={solid, fill=black, black}, forget plot]
  table[row sep=crcr]{%
1070.20833333333	17.4937816985149\\
2140.41666666667	19.5801442697261\\
4280.83333333333	21.4857281761791\\
8561.66666666667	23.1936047977965\\
17123.3333333333	25.2337458088959\\
};
\end{axis}
\begin{axis}[%
width=\Width,
height=\Height,
at={(0in,2.50in)},
scale only axis,
xmode=log,
xmin=200,
xmax=50000,
xtick={1000,10000},
xticklabels={{$\text{10}^\text{3}$},{$\text{10}^\text{4}$}},
xminorticks=true,
ylabel style={rotate=-90},
xlabel style={font=\color{white!15!black}},
separate axis lines,
ylabel={\color{mycolor1}$\overline{u'^2}^+$},
axis y line*= right,
yticklabel pos=right,
ylabel shift = -5pt,
title style={yshift=-5pt},
ytick pos=right,
separate axis lines,
every y tick label/.append style={font=\color{mycolor1}},
every y tick/.append style={mycolor1},
ymin=5,
ymax=9,
y axis line style  ={mycolor1},
ylabel style={font=\color{mycolor1}},
axis background/.style={fill=none}
]
\addplot [color=mycolor1, line width=1pt, forget plot]
  table[row sep=crcr]{%
1070.20833333333	8.56639572557865\\
17123.3083333333	6.05515569336378\\
};
\addplot [color=mycolor1, line width=0.0pt, draw=none, mark=square*, mark size=2pt, mark options={solid, fill=mycolor1, mycolor1}, forget plot]
  table[row sep=crcr]{%
1070.20833333333	8.45636727029621\\
2140.41666666667	8.01651837034099\\
4280.83333333333	7.50456254862524\\
8561.66666666667	6.5017045798506\\
17123.3333333333	6.0747224723026\\
};

\end{axis}

\begin{axis}[%
width=\Width,
height=\Height,
at={(0in,0.0in)},
scale only axis,
xmode=log,
xmin=200,
xmax=50000,
xtick={1000,10000},
xticklabels={{$\text{10}^\text{3}$},{$\text{10}^\text{4}$}},
xminorticks=true,
ylabel style={rotate=-90},
xlabel style={font=\color{white!15!black} \small},
xlabel={$z^{+}$},
separate axis lines,
ytick pos=left,
ylabel shift = -12pt,
every outer y axis line/.append style={black},
every y tick label/.append style={font=\color{black}},
every y tick/.append style={black},
ylabel={$\theta^+$},
title = {Temperature Profiles},
ymin=337,
ymax=345,
axis background/.style={fill=none}
]
\addplot [color=black, line width=1.5pt, forget plot]
  table[row sep=crcr]{%
1070.20833333333	339.019677468863\\
17123.3083333333	343.917407566908\\
};
\addplot [color=black, line width=1.5pt, draw=none, mark=triangle*, mark options={solid, rotate=180, fill=black, black}, forget plot]
  table[row sep=crcr]{%
1070.20833333333	338.98180221609\\
2140.41666666667	340.311094621451\\
4280.83333333333	341.443818337994\\
17123.3333333333	343.913026884852\\
};
\end{axis}
\begin{axis}[%
width=\Width,
height=\Height,
at={(0in,0.0in)},
scale only axis,
xmode=log,
xmin=200,
xmax=50000,
xtick={1000,10000},
xticklabels={{$\text{10}^\text{3}$},{$\text{10}^\text{4}$}},
xminorticks=true,
ylabel style={rotate=-90},
xlabel style={font=\color{white!15!black}},
separate axis lines,
axis y line*= right,
every y tick label/.append style={font=\color{mycolor2}},
every y tick/.append style={mycolor2},
ylabel style={font=\color{mycolor2}},
ylabel shift = -4pt,
title style={yshift=-5pt},
ytick pos=right,
ymin=5,
ymax=9,
y axis line style={mycolor2},
ylabel={$\overline{\theta'^2}^+$},
axis background/.style={fill=none},
yticklabel pos=right
]
\addplot [color=mycolor2, line width=1pt, forget plot]
  table[row sep=crcr]{%
1070.20833333333	8.61762267735331\\
17123.3083333333	6.02305412876164\\
};
\addplot [color=mycolor2, line width=0pt, draw=none, mark=square*, mark size=2pt, mark options={solid, fill=mycolor2, mycolor2}, forget plot]
  table[row sep=crcr]{%
1070.20833333333	8.36084391778956\\
2140.41666666667	8.10628168743831\\
4280.83333333333	7.62794300586202\\
17123.3333333333	5.83492474734073\\
};
\end{axis}
\end{tikzpicture}%
	\caption{Mean and variance profiles for velocity (top) and temperature (bottom) as a function of wall distance and normalized by inner units. Mean quantities ($U^+$ and $\theta^+$) are given in black, while variances for velocity ($\overline{u'^2}^+$) and temperature ($\overline{\theta'^2}^+$) are denoted in blue and orange, respectively.}
	\label{fig:log_region}
\end{figure}
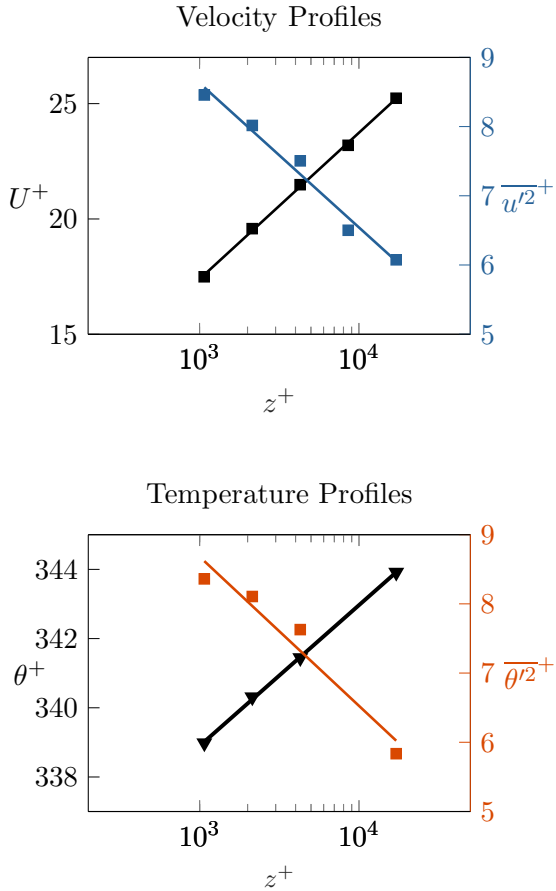

The existence of an inertial subrange can be substantiated by assessing whether the lower order statistics of the measurements, such as mean and variance, exhibit a canonical logarithmic behavior. Following \cite[][]{vallikivi2015turbulent},  who found logarithmic behavior above $z^+=zu_*/\nu>400$, the lowest measurement station to the ground at $z^+\approx$ 850 should be well within the log-layer. This assumption is further supported by the mean and variance profiles for the velocity and temperature shown in figure \ref{fig:log_region}, where logarithmic behavior is found in the region $10^3<z^+<10^4$. Here, $A_{1,u}$ and $A_{1,\theta}$ are estimated as 0.9 and 0.93, respectively, by a least squares fitting of the data to equation \ref{eq:gen_log_law_2}. This value for the velocity compares reasonably well the $A_1 \approx 1.25$ found by \cite{meneveau2013generalized}.  

\begin{figure*}[ht!]
	\centering
	\input{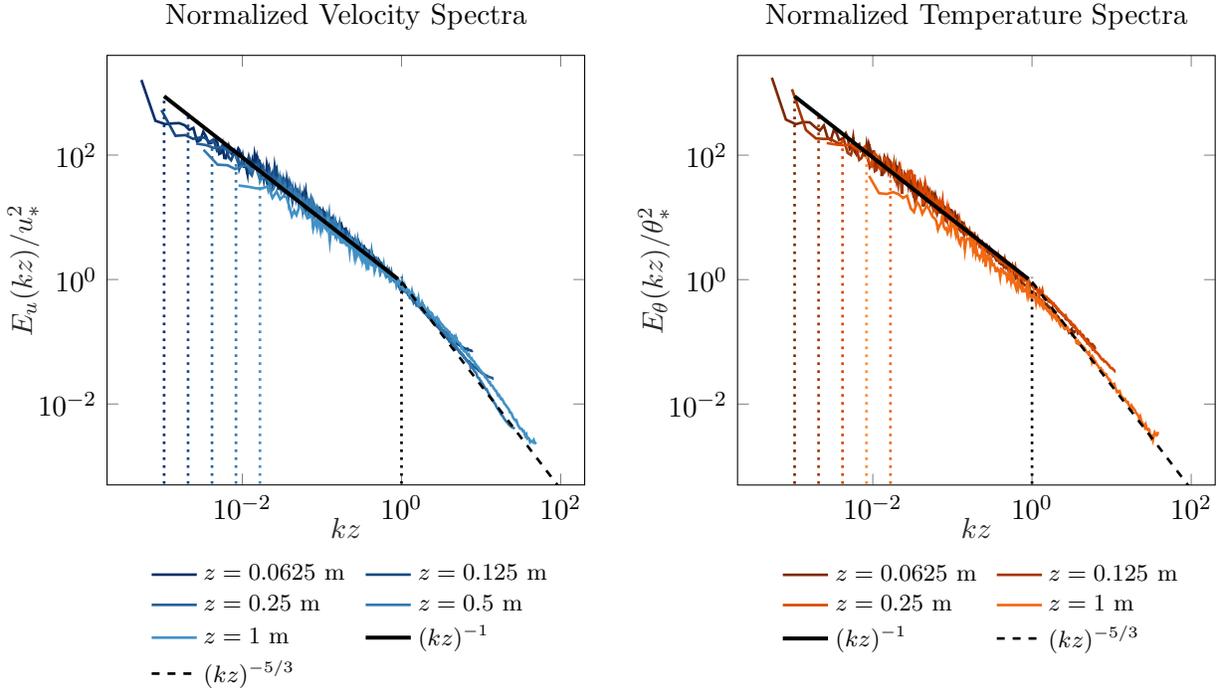}

	\caption{Normalized spectra of the longitudinal velocity ($E_u(k z)/u_*^2$) and temperature ($E_\theta(k z)/\theta_*^2$) are computed using the frozen turbulence hypothesis \citep{taylor1938spectrum}  and shown as a function of normalized wavenumber $k z$ in the left and right plots, respectively. Different shades of blue and orange indicate the respective spectra for velocity and temperature at the different measurement stations with lighter shades corresponding to distances further from the ground. For each quantity, there is a clear break point between the $k^{-1}$ (\myline{black}) and the $k^{-5/3}$ (\mydash{black}) scaling regimes as indicated by the black vertical dotted line (\thindot{black}) at $k=1/z$. The $k=1/\delta$ cutoff between the $k^{2}$ 
	and the $k^{-1}$ scaling regimes are also shown for each measurement height and denoted by their respective colors noted in the legend below each plot.}
	\label{spectra}
\end{figure*}

Finally, the streamwise wavenumber spectra of the velocity and temperature fluctuations can be compared to the idealized model, each of which are shown in figure \ref{spectra}. In both quantities, distinct $k^{-1}$ and $k^{-5/3}$ scaling regions are identified, consistent with the assumptions of the simplified model. However, due to the lack of convergence and stationarity of the flow, no data are available in the low wavenumber region where the  $k^2$ scaling is assumed. 
The expected area under the modeled $k^{2}$ region is identically equal to $\frac{1}{3}C_1$ in normalized variables. Compared to the integrated normalized spectrum of figure \ref{spectra} for each case, this comes out to between $3.90\%$ and $5.63\%$ of the total integrated area, whereas the $k^{-1}$ region represents $69.3\%$ to $79.6\%$  and the $k^{-5/3}$ region represents $16.5\%$ to $25.1\%$, as seen in Table \ref{tab:integrated area}. As the contribution from the $k^2$ region only affects the offset $B_{1,\theta}$ in equation \ref{eqn:HO_mom}, and the relative contribution remains small, the overall agreement between the model spectrum and the dataset is acceptable for the purposes here. Had a constant value for the spectrum been used such as \cite{katul2016generalized} or \cite{huang2021higher}, the area would have been significantly higher, overemphasizing the contribution from these wavenumbers and increasing the offset $B_{p,\theta}$.


\begin{table}
\centering
\small
    \begin{tabular}{cccc}
    \noalign{\smallskip}
  $z$  & $k^2$ & $k^{-1}$ & $k^{-5/3}$ \\
   $(\textrm{m})$& \multicolumn{3}{c}{(percentage contribution)}\\
    \noalign{\smallskip}\hline\hline\noalign{\smallskip}

    $0.0625$ & $3.9$ & $79.6$ & $16.5$ \\
    $0.125$ & $4.2$ & $77.6$ & $18.2$ \\
    $0.25$ & $4.6$ & $75.3$ & $20.1$ \\
    $1.0$ & $5.6$ & $69.3$ & $25.1$\\
    \noalign{\smallskip}\hline
    \end{tabular}
    \caption{Percentage of area represented by the three regions of the model spectrum for each measurement station.}
    \label{tab:integrated area}      
\end{table}

\section*{DISCUSSION}

Given the strength of the underlying assumptions and the similarity between the statistical behavior of temperature and velocity series, the predictive capability of the proposed RSDH model in capturing the behavior of the high-order moments can be assessed. Because the temperature exhibits a distribution that more closely follows Gaussian behavior, the RSDH predictions for high-moments should hold for $(\overline{\theta'^{2p}}^+)^{1/p}$. A comparison between the high-order moments and the RSHD predictions can be seen in figure \ref{figure6} for both the velocity and temperature measurements. In each case, the slope and offset of the $p=1$ statistics (i.e., the variances) are determined from integration of the spectral RSDH models given by \cite{katul2016generalized} and the present study (see equation \ref{eqn:sigma2}) rather than fitting a regression to the variance profiles. Similarly, the RSDH predictions for the high-order moments are determined from the $p=1$ case through equation \ref{eqn:HO_mom} that can be applied analogously to both velocity and temperature. 

Figure \ref{figure6} shows that for both temperature and velocity, there appears to be good agreement between the measurements and their respective spectral RSDH models. In each case, the model is well within the statistical uncertainty, with the exception of the $p=2$ case, partially due to the additive $(3/2) \overline{\theta'^{2p}}$ contribution as noted in the theory section of the manuscript. Further, it is difficult to discern any significant differences in the performance of the model between the two quantities.


\begin{figure}[ht!] 
\centering
	\input{figures/fig files/HO_mom2}
	\caption{Profiles of high-order moments of velocity (top) and temperature (bottom) as a function of normalized distance from the ground $z/\delta$. Central moments for $2p= 2,\,4,\,6,\, \mathrm{and}\;8$ are denoted by ($\circ$), ($\triangledown$), ($\lhd$), and ($\rhd$) symbols, respectively. The dashed lines represent the predictions for high-order moments as computed using equation \ref{eq:gen_log_law_2} for the velocity statistics \citep{katul2016generalized}  and equation \ref{eqn:HO_mom} for the temperature statistics. High-moments for velocity (blue) and temperature (orange) are denoted with darker shades of their respective colors. Error bars indicate errors due to limited statistical convergence of high-order moments.}
	\label{figure6}
\end{figure}

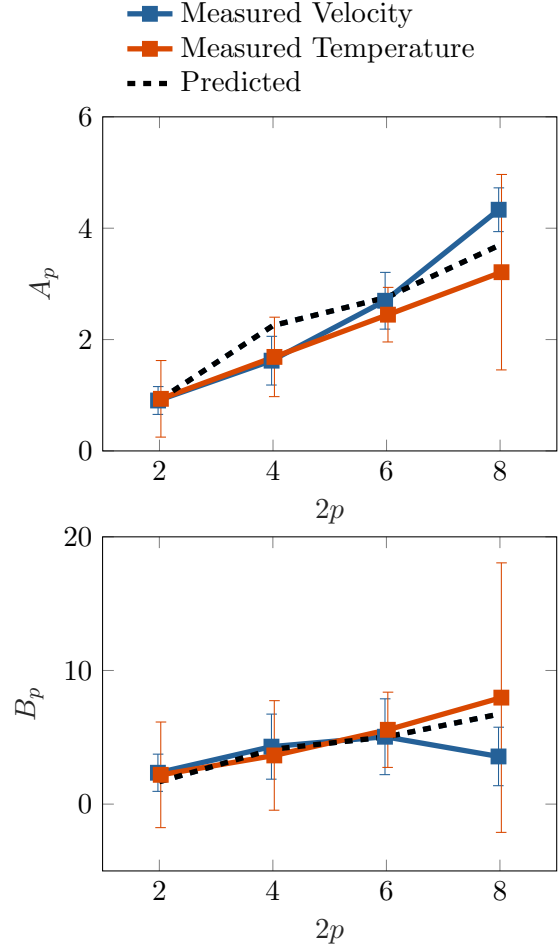
\begin{figure}[ht!] 
	\centering
%
%
\definecolor{mycolor1}{rgb}{0.14510,0.38039,0.59804}%
\definecolor{mycolor2}{rgb}{0.85098,0.28235,0.00392}%
\begin{tikzpicture}

\begin{axis}[%
width=2.377in,
height=1.75in,
at={(0in,2.2in)},
scale only axis,
xmin=1,
xmax=9,
xlabel style={font=\color{white!15!black}},
xlabel={$2p$},
ymin=0,
ymax=6,
ylabel style={font=\color{white!15!black}},
ylabel={$A_p$},
axis background/.style={fill=white},
legend style={at={(0.03,1.03)}, anchor=south west, legend cell align=left, align=left, draw=white}
]
\addplot [color=mycolor1, line width=2.0pt, mark size=2.0pt, mark=square*, mark options={solid, fill=mycolor1, mycolor1}]
 plot [error bars/.cd, y dir = both, y explicit]
 table[row sep=crcr, y error plus index=2, y error minus index=3]{%
1.975	0.905738862217024	0.250034869195757	0.250034869195757\\
3.975	1.62086899120303	0.437347807315772	0.437347807315772\\
5.975	2.696889672771	0.509624658789325	0.509624658789327\\
7.975	4.33163082969873	0.393047469162113	0.393047469162113\\
};
\addlegendentry{Measured Velocity}

\addplot [color=mycolor1, dashed, line width=2.0pt, forget plot]
  table[row sep=crcr]{%
2	0.91914834446279\\
4	2.25144444185774\\
6	2.75744503338837\\
8	3.69971215357186\\
};
\addplot [color=mycolor2, line width=2.0pt, mark size=2.0pt, mark=square*, mark options={solid, fill=mycolor2, mycolor2}]
 plot [error bars/.cd, y dir = both, y explicit]
 table[row sep=crcr, y error plus index=2, y error minus index=3]{%
2.025	0.93579328699844	0.687716024381171	0.687716024381171\\
4.025	1.68859178258103	0.713300055571163	0.713300055571161\\
6.025	2.4471098641349	0.490047978059641	0.490047978059643\\
8.025	3.2099875720366	1.75494546477128	1.75494546477128\\
};
\addlegendentry{Measured Temperature}

\addplot [color=black, dashed, line width=2.0pt]
  table[row sep=crcr]{%
2	0.91914834446279\\
4	2.25144444185774\\
6	2.75744503338837\\
8	3.69971215357186\\
};
\addlegendentry{Predicted}

\end{axis}

\begin{axis}[%
width=2.377in,
height=1.75in,
at={(0in,0in)},
scale only axis,
xmin=1,
xmax=9,
xlabel style={font=\color{white!15!black}},
xlabel={$2p$},
ymin=-5,
ymax=20,
ylabel style={font=\color{white!15!black}},
ylabel={$B_p$},
axis background/.style={fill=white}
]
\addplot [color=mycolor1, line width=2.0pt, mark size=2.0pt, mark=square*, mark options={solid, fill=mycolor1, mycolor1}, forget plot]
 plot [error bars/.cd, y dir = both, y explicit]
 table[row sep=crcr, y error plus index=2, y error minus index=3]{%
1.975	2.34674738569717	1.39209726509521	1.39209726509521\\
3.975	4.3006003137524	2.43498312222608	2.43498312222607\\
5.975	5.04214718331227	2.83739262450735	2.83739262450734\\
7.975	3.56658835484089	2.18833600542646	2.18833600542647\\
};
\addplot [color=mycolor1, dashed, line width=2.0pt, forget plot]
  table[row sep=crcr]{%
2	1.67606132266311\\
4	4.10549501813889\\
6	5.02818396798932\\
8	6.74640223522641\\
};
\addplot [color=mycolor2, line width=2.0pt, mark size=2.0pt, mark=square*, mark options={solid, fill=mycolor2, mycolor2}, forget plot]
 plot [error bars/.cd, y dir = both, y explicit]
 table[row sep=crcr, y error plus index=2, y error minus index=3]{%
2.025	2.19159260675539	3.95169712307551	3.95169712307551\\
4.025	3.64241926805177	4.09870597391788	4.09870597391789\\
6.025	5.56124289475028	2.81587329131934	2.81587329131933\\
8.025	7.96883867415368	10.0841229496309	10.0841229496309\\
};
\addplot [color=black, dashed, line width=2.0pt, forget plot]
  table[row sep=crcr]{%
2	1.67606132266311\\
4	4.10549501813889\\
6	5.02818396798932\\
8	6.74640223522641\\
};
\end{axis}
\end{tikzpicture}%
	\caption{Coefficients $A_p$ (blue) and $A_{p,\theta}$ (orange) as a function of moment order $2p$, averaged across available datasets. Square markers ($\square$) denote slopes ($A_p$) and offsets ($B_p$) determined through a linear regression of the central moment profiles for $2p= 2,\,4,\,6,\, \mathrm{and}\;8$. Error bars represent a $95\%$ confidence interval on the regression coefficients from fitting estimator values. The dashed lines represent the coefficients as predicted by integration of the spectral model and RSDH predictions given by equations \ref{eq:gen_log_law_2} and \ref{eqn:HO_mom}.  Due to their similarity, these RSDH predictions appear overlapping.}
	\label{figure7}
\end{figure}

However, the differences between the temperature and velocity statistics become more apparent in figure \ref{figure7}, which directly compares the slopes ($A_p$) and offsets ($B_p$) obtained from the measurements and RSDH model. There, the slope ($A_p$) and offset ($B_p$) values from the RSDH predictions are extracted from the lines shown in figure \ref{figure6} and plotted in the left and right plots of figure \ref{figure7}, respectively, as a function of moment order $p$. These predictions are compared with experimental values obtained from a linear regression of the high order moments with the logarithmic wall distance, with error bars representing a $95\%$ confidence interval on these calculated values. The performance of the RSDH model is comparable for velocity and temperature for $p\leq3$, with the greatest deviation appearing at $p=4$. The temperature prediction all falls within the $95\%$ confidence interval for each value, while velocity has a small deviation for the even moments, albeit with narrower confidence intervals. This discrepancy is expected to widen for progressively larger values of $p$ due to the increasing prevalence of the distribution tails to the higher order moments; however, such high-order moments are not analyzed here due to limited convergence of these statistics in the present dataset.

\begin{figure*}[ht!]
	\centering
	\input{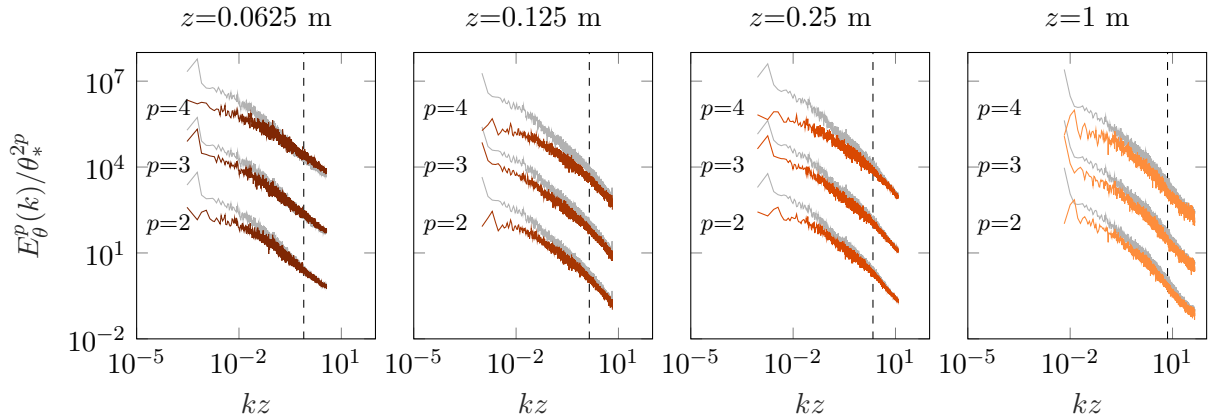}
	\caption{Calculated higher order spectra ($p\geq2$), shown in color, compared with the theoretical spectra from the RSDH, shown in grey, for each measurement location. Dashed black lines indicate $k z = z / \lambda$, where $\lambda$ is the Taylor microscale}
	\label{figure8}
\end{figure*}

Discrepancies that arise in the theoretical results can be explained by the nature of equation \ref{eqn:HO_spectra}, which was derived by \cite{van1975higher} under the assumption that the higher wavenumbers in the spectrum $E(k)$ followed a $k^{-5/3}$ model proposed by von K\'{a}rm\'{a}n and discussed by \cite{hinze1959turbulence}. The integration of this model assumed large wavenumbers and considered only the leading term in a convolution following an asymptotic expansion. The resulting expression is equation \ref{eqn:HO_spectra}, which has explicitly not treated the low wavenumbers, including the $k^{-1}$ region. As this model breaks down for wavenumbers on the order of the Taylor microscale, the lower wavenumber regions should depart from the theoretical inertial range behavior. This deviation is evident in figure \ref{figure8}, which compares the actual higher order temperature spectrum ($p\geq2$, in color) with the theoretical spectrum from the RSDH (shown in grey). The deviation between actual and calculated can be seen to be exacerbated for the lower wavenumber region, with the location of the Taylor microscale indicated by the vertical dashed lines. This results in the RSDH over-predicting the higher order spectrum throughout the datasets. This behavior is consistent with \cite{huang2021higher}, which demonstrated the RSDH over-predicted the higher order velocity spectra.  However, the hard truncation of the spectral model at the Batchelor scale, the assumed Saffman spectrum in the lowest $k$ region, and the piecewise discontinuities between each region will likely mean a loss in energy from the RSDH, which may compensate for the over-prediction in the large wavenumbers.

Another potential cause of deviations from the RSDH stems from the assumption of neutral stability conditions. While the Monin-Obukhov stability parameter and flux Richardson numbers in this study both indicate near-neutral conditions, the effect of buoyancy within the flow field will distort the slope of a logarithmic region \citep{cheng2021logarithmic}. The lack of universality when non-neutral conditions arise could be tied to the skewing of the PDFs from Gaussian behavior, which alters the parameter $\alpha(p)$ and therefore the constants in figure \ref{figure7}.

\section*{CONCLUSIONS}

The similarity between temperature and momentum is used to propose a new theory for the behavior of passive scalar moments in the inertial sublayer based on RSDH. The predicted behavior from the new model is found to be in good agreement with highly resolved measurements of temperature obtained in the atmospheric surface layer under near-neutral conditions with deviations smaller from RSDH behavior than their collocated streamwise velocity counterpart. One potential explanation for the higher deviation in the velocity moments is the non-locality of pressure acting on the velocity components, which is not present in the temperature field. Additionally, the ideal spectrum selected is continuous but not smooth and has a hard transition from inertial to viscous wavenumbers at the Batchelor scale. These, coupled with the multiplicative factor of $\alpha(p)$ in equation \ref{eqn:HO_spectra}, can combine to have competing effects of under-predicting and over-predicting the spectral energy, respectively. There is also some uncertainty in the estimate of the boundary layer height, which is estimated by comparing the $\overline{u'^2}$ profiles with the relation proposed by \citep{marusic2003streamwise}. This would contribute to a shift in the values of the integrated parameters, namely the relative amount of energy contributing to $A_{1,\theta}$ and $B_{1,\theta}$. While the performance of the new model is encouraging, extending and evaluating this framework under non-neutral stability conditions and turbulent $Pr \neq 1$ remains an area of future investigation.

\section*{ACKNOWLEDGMENTS}
GGK acknowledges support from the U.S. National Science Foundation (NSF-AGS-2028633) and the Department of Energy (DE-SC0022072), KYH from ONR Grant No. N00014-21-1-2296 (Fatima Multidisciplinary University Research Initiative) administered by the Marine Meteorology and Space Program of the Office of Naval Research, and ADB from the U.S. National Science Foundation (NSF-CBET-2042346).

\appendix

\section{Boundary Layer Parameters}
\label{sec:appendix}

The number of eddy turnovers in the datasets are calculated in two different ways. An estimate of the number large scale eddy turnovers, $N_e = T_{test}/(\delta/\bar{u})$, with $T_{test}$ the total dataset time, $\delta$ the boundary layer height, and $\bar{u}$ the mean velocity, for each of the four main datasets ranges from $196\leq N_e \leq 226$ for the four datasets. The mean velocity is taken at $z=2$ m, which provides a significant underestimate of the large scale eddy turnover time.

The local calculation for eddy turnovers, $N_{le}$, at each measurement station height is shown in table \ref{tab:Eddy_turnovers}, with $N_{le}= T_{test}/\mathcal{L}_T$ and $\mathcal{L}_T$ is the integral time scale calculated from the autocorrelation of the velocity data. This shows that each station has at least $500$ local eddy turnovers.
\begin{table}
\centering
\small
    \begin{tabular}{cc}
    \noalign{\smallskip}
  $z$   & $T_{test}/\mathcal{L}_T$ \\
  $(\textrm{m})$ & \\
    \noalign{\smallskip}\hline\hline\noalign{\smallskip}

     $0.0625$ & 739.6   \\
     $0.125$ & 1129 \\
    $0.25$  & 1180 \\
     $1.0$ & 560.4\\
    \noalign{\smallskip}\hline
    \end{tabular}
    \caption{Estimate of eddy turnover times for each sensor location. }
    \label{tab:Eddy_turnovers}      
\end{table}




\bibliographystyle{elsarticle-harv} 
\bibliography{tsfp}


\end{document}